\documentclass{aa}
\usepackage{graphicx}
\usepackage[varg]{txfonts}
\usepackage{natbib}
\bibpunct{(}{)}{;}{a}{}{,}
\usepackage{array}
\usepackage{xspace}
\usepackage{lscape}
\usepackage{url}
\usepackage{multirow}

\usepackage[hyperfootnotes=false, linktocpage=true, breaklinks=true, colorlinks=true, linkcolor=blue, citecolor=blue, urlcolor=blue]{hyperref}
\usepackage[all]{hypcap}

\usepackage{mathtools}
\usepackage[figuresright]{rotating}

\newcommand{\nni}{\ion{N}{i}\xspace}

\newcommand{\oi}{\ion{O}{i}\xspace}

\newcommand{\ovii}{\ion{O}{vii}\xspace}
\newcommand{\oviii}{\ion{O}{viii}\xspace}
\newcommand{\neix}{\ion{Ne}{ix}\xspace}
\newcommand{\nex}{\ion{Ne}{x}\xspace}
\newcommand{\mgxi}{\ion{Mg}{xi}\xspace}
\newcommand{\fexvii}{\ion{Fe}{xvii}\xspace}

\begin{document} 

\title{{\it Suzaku} and {\it XMM-Newton} Observations of the North Polar Spur: Charge Exchange or ISM Absorption?}

\author {Liyi Gu \inst{1}
\and 
Junjie Mao \inst{1}
\and
Elisa Costantini \inst{1}
\and
Jelle Kaastra \inst{1,2} }

\institute{
  SRON Netherlands Institute for Space Research, Sorbonnelaan 2, 3584 CA Utrecht, The Netherlands \\ \email{L.Gu@sron.nl}
\and 
Leiden Observatory, Leiden University, PO Box 9513, 2300 RA Leiden, The Netherlands 
}
%\and 
%Leiden Observatory, Leiden University, PO Box 9513, 2300 RA Leiden, The Netherlands  }
%\institute{SRON Netherlands Institute for Space Research, Sorbonnelaan 2, 3584 CA Utrecht, The Netherlands}

%\date{Received 15 April 2015 / Accepted 22 May 2015}

%==============%
\abstract
%==============%
{ 
  By revisiting the {\it Suzaku} and {\it XMM-Newton} data of the North Polar Spur, we discovered that the spectra are inconsistent
  with the traditional model consisting of pure thermal emission and neutral absorption. The most prominent discrepancies are the enhanced \ovii and \neix forbidden-to-resonance
  ratios, and a high \oviii Ly$\beta$ line relative to other Lyman series. A collisionally ionized absorption model can naturally explain both features, while
  a charge exchange component can only account for the former. By including the additional ionized absorption, the plasma in the North Polar
  Spur can be described by a single-phase CIE component with temperature of 0.25 keV, and nitrogen, oxygen, neon, magnesium, and iron abundances of $0.4-0.8$ solar. The abundance pattern
  of the North Polar Spur is well in line with those of the Galactic halo stars. The high nitrogen-to-oxygen ratio
  reported in previous studies can be migrated to the large transmission of the \oviii Ly$\alpha$ line. The ionized absorber is characterized by a balance temperature of
  $0.17-0.20$ keV and a column density of $3-5 \times 10^{19}$ cm$^{-2}$. Based on the derived abundances and absorption, we speculate that the
  North Polar Spur is a structure in the Galactic halo, so that the emission is mostly absorbed by Galactic ISM in the line of sight.
}


\authorrunning{L. Gu et al.}
\titlerunning{{\it Suzaku} and {\it XMM-Newton} Observations of the North Polar Spur}

\maketitle

%====================%
\section{Introduction}
%====================%

The North Polar Spur (NPS hereafter) is a prominent structure emitting both in the soft X-ray and radio bands,
with a projected distribution from the Galactic plane at $l \sim 20^{\circ}$ towards the north Galactic pole.
Despite its vicinity, the origin of the NPS remains largely unclear. Early research suggested that 
the NPS is an old supernova remnant, or a front created by stellar wind from the Scorpio-Centaurus OB association,
at a distance of several hundred pc from the Sun (Berkhuijsen et al. 1971; Egger \& Aschenbach 1995). 
Alternatively, the NPS can be explained as a shock front produced by an energetic event, such as starburst, 
in the Galactic center $\sim 15$ Myr ago (Sofue et al. 1977; Bland-Hawthorn \& Cohen 2003). Recent morphological
studies further indicated a possible relation between the NPS and the {\it Fermi} $\gamma-$ray bubbles (e.g., 
Kataoka et al. 2013). In the ``Galactic center origin'' scenario, the distance to the NPS is expected to be several kpc.

X-ray studies of the NPS hot plasma provide important information, including the plasma temperature,
density, and metal abundances, which are essential to understand its origin. Using {\it XMM-Newton} observations
of three regions in the NPS, Willingale et al. (2003, hereafter W03) identified a thermal component, with a temperature
of $\approx 0.26$ keV and metal abundances of $\sim 0.5$ $Z_{\odot}$, associated with the enhanced NPS emission.
They further deduced that the thermal energy contained in the NPS is consistent with the energy released by one or
more supernovae events. Based on a {\it Suzaku} observation, Miller et al. (2008, hereafter M08) measured a
slightly higher thermal temperature $\sim 0.30$ keV, and a quite enhanced nitrogen abundance, with nitrogen-to-oxygen
abundance ratio $\approx 4.0$ times of the solar value. They proposed that additional enrichment from stellar
evolution in the NPS vicinity is required to explain the observed abundance pattern.

The above two X-ray studies are both based on an assumption that the NPS emission is purely from thermal plasma in collisional
ionization equilibrium (CIE), affected by only neutral absorption in the line-of-sight. Recent research shows that
some diffuse objects also emit non-thermal X-rays, such as charge exchange emission produced at the interface between
hot and cold materials (e.g., Lisse et al. 1996, Katsuda et al. 2011, Gu et al. 2015). There are also cases in which a portion 
of the foreground absorber appears to be highly ionized, as reported in e.g., Yao \& Wang (2005) and Hagihara et al. (2010, 2011). Both the charge exchange component
and ionized absorption can strongly affect the line emission, and hence deviate the resulting physical model.
Indeed, the NPS is a potential target for charge exchange, because it might be surrounded by
a shell of neutral gas (e.g., Heiles et al. 1980), providing substantial environment for ion-neutral charge exchange. 
Meanwhile, the NPS might be subject to ionized absorption, since in the ``Galactic center origin'' scenario, it is 
expected to be located behind
a layer of hot interstellar medium (ISM) of the Galactic halo/bulge. Actually, the spectra presented in W03 and M08 
already showed a hint for such additional components: the central line energies of unresolved \ovii and \neix triplets 
are shifted by a few 10 eV to longer wavelength, relative to the energies of \oviii and \nex (Lallement 2009). This might indicate
either charge exchange enhancement of the forbidden lines, or possible absorption in the resonance ones.

In this paper, we present a detailed spectral analysis by revisiting the high quality X-ray data of the NPS.
The latest {\tt SPEX} version 3.01 is employed, as it includes a new charge exchange code (Gu et al. 2016) and
a model for ionized absorption. \S2 gives a brief description of data reduction, and the data analysis and results
are presented in \S3. We discuss the physical implication of the results in \S4 and summarize our work in \S5.
Throughout the paper, the errors are given at 68\% confidence level. We adopt the proto-Solar abundance table of
Lodders et al. (2009), and convert the previous abundance measurements to the new standard.

%====================%
\section{Observations and data reduction}
%====================%

%=========================%
\subsection{{\it Suzaku} and {\it XMM-Newton} datasets}
%=========================%

The NPS region was observed by {\it Suzaku}, pointing at Galactic coordinates $l=26.83^{\circ}$ and $b=21.95^{\circ}$, on 2005 October 3 for a total exposure of 46.1 ks. The same XIS dataset
was already utilized in M08. The data were processed with the latest HEASoft 6.18 and CALDB 151005. 
Following Gu et al. (2012), we removed hot pixels and data obtained either near South Atlantic Anomaly or at low elevation
angles from the Earth rim. For each XIS detector, a $0.3-8.0$ keV lightcurve was extracted from a source free region,
and was screened to filter off anomalously high count bins with rates above the 2$\sigma$ limit of the quiescent mean value.
The XIS0 and XIS1 data are affected by high count rate periods over 3$\sigma$ of the mean value, probably due to temporary
  system anomaly and variation in the particle background. Some high particle periods are also visible in the XIS2 and XIS3 lightcurves. 
The clean exposures are 37 ks, 38 ks, 40 ks, and 40 ks for the XIS0, XIS1, XIS2, and XIS3 data, respectively.
Similar time filtering results were reported in M08. Contaminating point sources were masked out in the same way as M08.

{\it XMM-Newton} was used to observe six fields of the NPS region; as reported in W03, only three datasets (fields IV, V, and VI) are useful, providing a total exposure of 45.7 ks.
The three pointings have Galactic ($l,b$) = ($25^{\circ}, 20^{\circ}$), ($20^{\circ}, 30^{\circ}$), and ($20^{\circ}, 40^{\circ}$).
The SAS v13.5 and the built-in extended source analysis software (ESAS) were
utilized to process and calibrate the data
obtained with the {\it XMM-Newton} European Photon Imaging Camera (EPIC). The MOS raw data were created by {\tt emchain},
and the lightcurves were extracted and screened for time variable background component by the {\tt mos-filter} task, which uses
filter of 2$\sigma$ as a conservative cut. The final net clean
exposures are 13 ks, 12 ks, and 13 ks for fields IV, V, and VI, respectively. The point sources were detected and removed
by {\tt cheese} task with a flux threshold of $10^{-14}$ ergs cm$^{-2}$ s$^{-1}$; the field IV is also contaminated by a diffuse
X-ray source and it was masked out manually. The spectra and response files were calculated by the {\tt mos-spectra} mask.
The pn data are not included in this work, since they suffered more from the time variable particle background, and have a lower
spectral resolution than the MOS data in soft X-ray band.

%=========================%
\subsection{Background modeling}
%=========================%

The background was estimated as a combination of three components, i.e., non X-ray background (NXB), cosmic X-ray background
(CXB), and Galactic emission. For the XIS and MOS datasets, the NXB spectra were created by the {\tt xisnxbgen} and {\tt mos-back} tasks,
which are based on a dark earth observation and exposures with filter wheel closed, respectively. To determine the CXB and Galactic
  components, we analyzed an off-source XIS pointing to the direction of an intermediate polar 1RXS J180340.0+40121 at Galactic $l = 66.85^{\circ}$
  and $b = 25.78^{\circ}$. This object is clearly outside the NPS structure, and has a similar Galactic latitude as the NPS XIS pointing.
  Hence, it is optimal for characterizing the background of the NPS structure.
  The data was screened in the same way as described in \S2.1. After excluding the central $4^{\prime}$ region
  covering the intermediate polar, and subtracting the NXB based on dark earth observation, we perform a model fitting of the remaining
spectrum with the CXB and Galactic components. The CXB is
approximated by a broken powerlaw with photon indices $\Gamma =$ 2.0 and 1.4 at $<0.7$ keV and $>0.7$ keV bands, respectively,
absorbed by the Galactic cold material with column density of $3.54 \times 10^{20}$ cm$^{2}$ (Willingale et al. 2013). The Galactic
emission consists of two sub-components, i.e., local hot bubble (LHB) and Galactic halo (GH) emissions. We include an unabsorbed,
solar-abundance {\tt CIE} component with a fixed temperature = 0.08 keV to account the LHB emission, and another absorbed brighter
{\tt CIE} component with temperature of 0.2 keV and solar-abundance for the GH. The off-source XIS spectra can be well fit by the model,
with C-statistics of 215 for 182 degrees of freedom. The best-fit CXB flux is 7.2 $\times 10^{-8}$ ergs cm$^{-2}$ s$^{-1}$ sr$^{-1}$ in
$2-10$ keV, which agrees well with previous reportings (e.g., Gu et al. 2012). The best-fit CXB and Galactic backgrounds are then used
in the subsequent NPS analysis.

Errors quoted in the subsequent spectral analysis were estimated by accounting for both statistical and systematic uncertainties.
The former were calculated by the {\tt SPEX} command {\tt error}, as the fitting was repeated for several iterations to ensure
that the actual minimum C-statistics is found. For the latter, the NXB, CXB, LHB and GH components were re-normalized by 10\% to assess
the typical background error.

%====================%
\section{Analysis and results}
%====================%

%=========================%
\subsection{{\tt CIE}/{\tt CX} with neutral absorption}
%=========================%

%============================
%  FIG: neu-CIE
%
\begin{figure*}[!]
\centering
\resizebox{0.8\hsize}{!}{\hspace{-1cm}\includegraphics[angle=0]{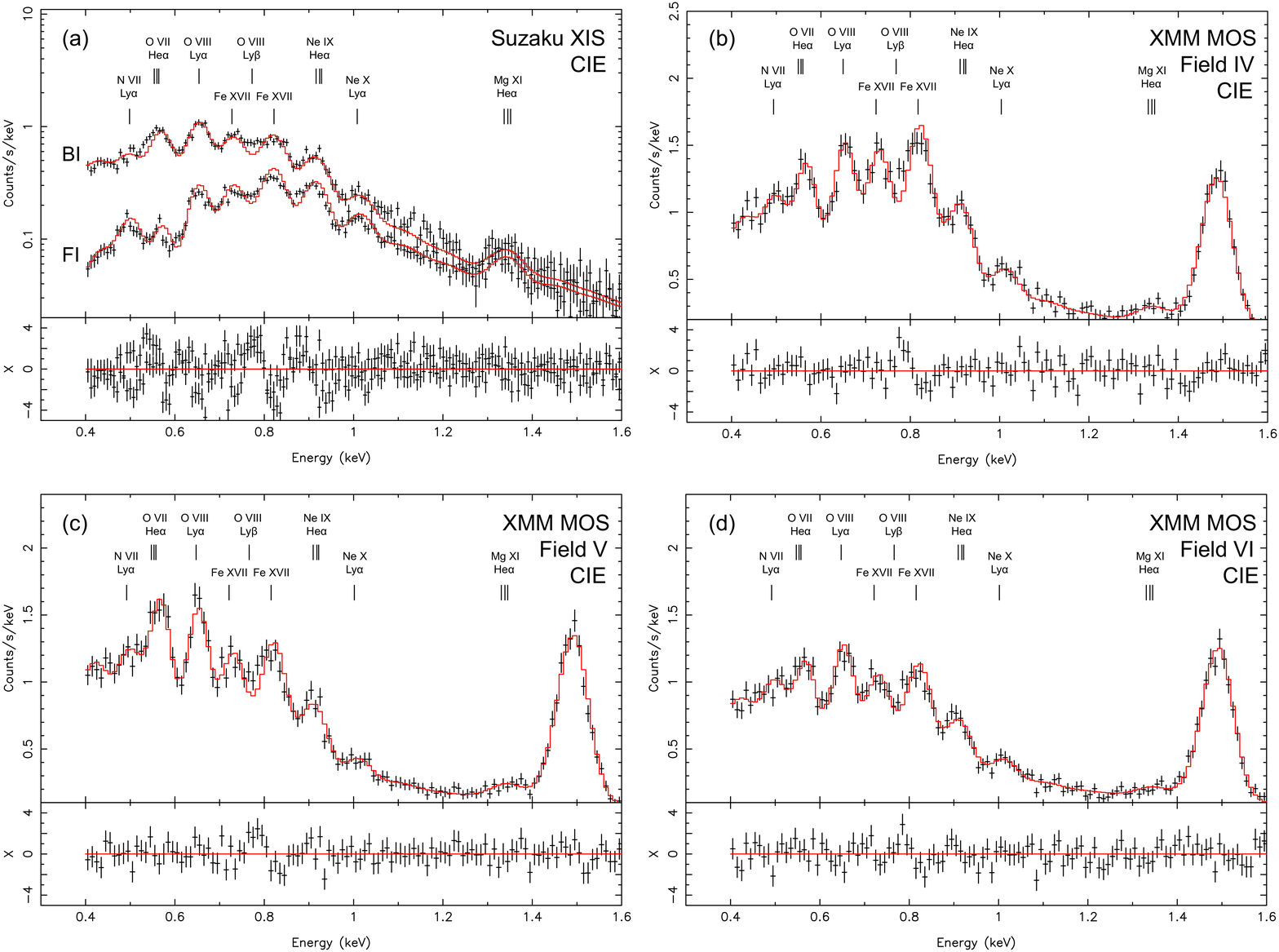}}
\caption{The NPS spectra modeled with a single-phase {\tt CIE} component absorbed by neutral material. The fittings of the {\it Suzaku}
  XIS are shown in (a), and {\it XMM-Newton} MOS field IV, V, and VI data are in panels (b), (c), and (d), respectively. Strong emission lines are marked in text. The line
  at $\approx 1.5$ keV in each MOS spectrum is an instrumental line. 
\vspace{0.5cm}
}
\label{CIE_fig}
\end{figure*}
%============================

%============================
%  FIG: neu-CIE
%
\begin{figure*}[!]
\centering
\resizebox{0.8\hsize}{!}{\hspace{-1cm}\includegraphics[angle=0]{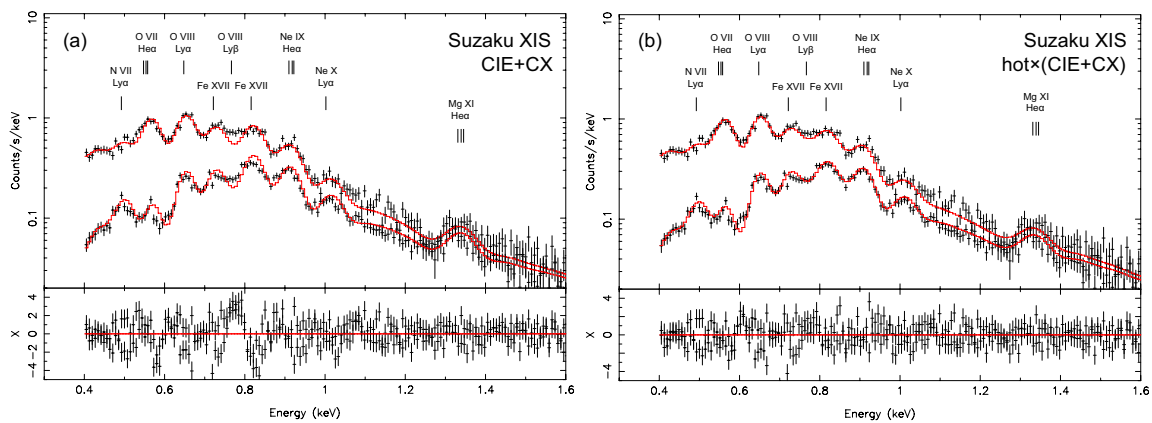}}
\caption{The {\it Suzaku} NPS spectra modeled with {\tt CIE} + {\tt CX} components absorbed by neutral material (a), and absorbed by both neutral and ionized materials (b).
  Strong emission lines are marked in text. 
\vspace{0.5cm}
}
\label{CIE_fig}
\end{figure*}
%============================

%============================
%  TABLE: best-fit parameters
%
\begin{table*}[!]
\begin{minipage}[t]{\hsize}
\setlength{\extrarowheight}{3pt}
\caption{Best-fit parameters of the NPS spectra}
\label{lines_table}
\centering
\small
\renewcommand{\footnoterule}{}
\begin{tabular}{l l c c | c c}
  \hline \hline
  & & \multicolumn{2}{c|}{Neutral absorber} & \multicolumn{2}{l}{Ionized + neutral absorbers} \\
  & & {\tt CIE} & {\tt CIE} + {\tt CX} & {\tt CIE} & {\tt CIE} + {\tt CX} \\  
  \hline
  \multirow{9}{*}{{\it Suzaku} XIS} & $k$T\tablefootmark{a} (keV)      & $0.27\pm0.01$ & $0.28\pm0.01$ & $0.25\pm0.01$ & $0.25\pm0.01$\\
                     & N ($Z_{\odot}$)  & $0.71\pm0.17$          & $0.59\pm0.12$          & $0.48\pm0.15$          & $0.49\pm0.17$          \\
                     & O ($Z_{\odot}$)  & $0.25\pm0.03$          & $0.21\pm0.04$          & $0.71\pm0.09$          & $0.65\pm0.11$          \\
                     & Ne ($Z_{\odot}$) & $0.32\pm0.03$          & $0.23\pm0.03$          & $0.48\pm0.06$          & $0.44\pm0.07$          \\
                     & Mg ($Z_{\odot}$) & $0.38\pm0.06$          & $0.25\pm0.05$          & $0.46\pm0.08$          & $0.43\pm0.08$          \\
                     & Fe ($Z_{\odot}$) & $0.54\pm0.06$          & $0.45\pm0.04$          & $0.65\pm0.07$          & $0.63\pm0.07$          \\
		     & $k$T$_{\it hot}$\tablefootmark{b} (keV)         & $-$                    & $-$                   & $0.19\pm0.01$ & $0.20\pm0.02$\\
		     & nH$_{\it hot}$\tablefootmark{c} ($10^{19}$ cm$^{-2}$)     & $-$         & $-$                    & $3.8\pm0.3$    & $5.7\pm1.3$   \\ 
                     & C-stat/dof      & 1152.9/717           & 1081.7/715           & 902.3/714    & 877.2/712 \\
  \hline
  \multirow{8}{*}{{\it XMM} MOS field IV} & $k$T\tablefootmark{a} (keV)   & $0.26\pm0.01$ & $-$ & $0.24\pm0.01$ & $-$   \\
                     & O ($Z_{\odot}$)  & $0.28\pm0.05$          & $-$                    & $0.68\pm0.07$         & $-$ \\
                     & Ne ($Z_{\odot}$) & $0.35\pm0.08$          & $-$                    & $0.54\pm0.07$         & $-$ \\
                     & Mg ($Z_{\odot}$) & $0.70\pm0.22$          & $-$                    & $0.78\pm0.24$         & $-$ \\
                     & Fe ($Z_{\odot}$) & $0.72\pm0.07$          & $-$                    & $0.74\pm0.09$         & $-$ \\
		     & $k$T$_{\it hot}$\tablefootmark{b} (keV)         & $-$                    & $-$                    & $0.18\pm0.01$ & $-$ \\
		     & nH$_{\it hot}$\tablefootmark{c} ($10^{19}$ cm$^{-2}$)  & $-$            & $-$                    & $3.2\pm0.3$    &  $-$ \\ 
                     & C-stat/dof      & 148.8/111            & $-$                   & 130.5/108   & $-$ \\

  \hline
  \multirow{8}{*}{{\it XMM} MOS field V} & $k$T\tablefootmark{a} (keV)   & $0.24\pm0.01$ & $-$ & $0.23\pm0.01$ & $-$\\
                     & O ($Z_{\odot}$)  & $0.27\pm0.06$          & $-$                   & $0.62\pm0.08$       & $-$ \\
                     & Ne ($Z_{\odot}$) & $0.25\pm0.05$          & $-$                   & $0.42\pm0.07$       & $-$ \\
                     & Mg ($Z_{\odot}$) & $0.62\pm0.19$          & $-$                   & $0.78\pm0.27$       & $-$ \\
                     & Fe ($Z_{\odot}$) & $0.54\pm0.06$          & $-$                   & $0.59\pm0.06$       & $-$ \\
		     & $k$T$_{\it hot}$\tablefootmark{b} (keV)         & $-$                    & $-$                    & $0.19\pm0.01$ & $-$ \\
		     & nH$_{\it hot}$\tablefootmark{c} ($10^{19}$ cm$^{-2}$)     & $-$         & $-$                    & $2.7\pm0.4$     & $-$ \\ 
                     & C-stat/dof    & 130.4/109            & $-$                    & 104.3/106   & $-$ \\
  \hline
  \multirow{8}{*}{{\it XMM} MOS field VI} & $k$T\tablefootmark{a} (keV)  & $0.25\pm0.01$ & $-$ & $0.23\pm0.01$ & $-$\\
                     & O ($Z_{\odot}$)  & $0.23\pm0.06$          & $-$                   & $0.68\pm0.21$          & $-$ \\
                     & Ne ($Z_{\odot}$) & $0.26\pm0.07$          & $-$                   & $0.61\pm0.17$          & $-$ \\
                     & Mg ($Z_{\odot}$) & $0.49\pm0.16$          & $-$                   & $0.68\pm0.29$          & $-$ \\
                     & Fe ($Z_{\odot}$) & $0.56\pm0.09$          & $-$                   & $0.75\pm0.12$          & $-$ \\
		     & $k$T$_{\it hot}$\tablefootmark{b} (keV)         & $-$                    & $-$                    & $0.17\pm0.02$ & $-$ \\
		     & nH$_{\it hot}$\tablefootmark{c} ($10^{19}$ cm$^{-2}$)     & $-$         & $-$                    &  $2.8\pm0.9   $ & $-$ \\ 
                     & C-stat/dof    & 126.1/109           & $-$                     & 115.7/106  & $-$\\
\hline
\hline

\end{tabular}
\end{minipage}
 \tablefoot{
   \tablefoottext{a}{Best-fit plasma temperature of the {\tt CIE} component.} \\
   \tablefoottext{b}{Best-fit plasma temperature of the ionized absorber.} \\
   \tablefoottext{c}{Best-fit column density of the ionized absorber.}   
}
\end{table*}

After subtracting the point sources and background, we first fit the full-field XIS and MOS spectra by a single-phase {\tt CIE} component,
absorbed by only neutral material. For each field, the absorbing column density is allowed to vary up to the Galactic value given in Willingale et al. (2013). The plasma emission
measure, temperature, and N, O, Ne, Mg, and Fe abundances are also left free. The fitting was performed over the $0.4-3.0$ keV
for both XIS and MOS data. The best-fit models are plotted in Figure 1. As was already noted in M08, the XIS spectra in the
$0.5-1.0$ keV are poorly fit with this simple model (C-statistics = 1152.9 for a degree of freedom of 717). Apparent residuals are seen
in the \ovii He$\alpha$ ($0.56-0.57$ keV), \neix He$\alpha$ ($0.90-0.92$ keV) and \oviii Ly$\beta$ (0.77 keV) bands. At the first two
energies, the emission from the unresolved He-like triplets appears to be systematically shifted, by $10-20$ eV, to the forbidden line side,
while the H-like Ly$\alpha$ counterparts are nicely fitted by the thermal model. For the \oviii Ly$\beta$ line, the {\tt CIE} model
underestimates the line intensity by $\sim 20$\%, which indicates an anomalously high Ly$\beta$/Ly$\alpha$ ratio. Adding another {\tt CIE}
component does not improve the fitting. The poor fit cannnot be caused by possible defects in the {\it Suzaku} calibration or atomic model (Appendix A).
As shown in Table 1, the best-fit temperature is $0.27 \pm 0.01$ keV, the N and O abundances are $0.71\pm0.17$ $Z_{\odot}$ and $0.25\pm0.03$ $Z_{\odot}$, respectively,
and the Ne, Mg, and Fe abundances are about $0.3-0.6$ $Z_{\odot}$. These results agree well with those reported in M08.

As plotted in Figure 1b$-$1d, the MOS spectra are fit with the CIE model with neutral absorber, and the obtained C-statistics are 148.8, 130.4, and 126.1 for $\approx 110$ degrees of freedom for field IV, V, and VI, respectively. Compared to the XIS data, the MOS spectra are better fit by the model in the
the \ovii He$\alpha$ and \neix He$\alpha$ bands, while apparent residuals can still be seen at \oviii Ly$\beta$. The resulting temperatures
and metal abundances are presented in Table 1. Similar to the XIS results, the three MOS fields give temperatures of $0.24-0.26$ keV, O abundances of $\sim 0.2-0.3$ $Z_{\odot}$, and Ne, Mg, and Fe values of $0.2-0.7$ $Z_{\odot}$. The best-fit parameters are consistent with those reported in W03.

%============================
%  FIG: lbla
%
\begin{figure*}[!]
\centering
\resizebox{0.8\hsize}{!}{\hspace{-1cm}\includegraphics[angle=0]{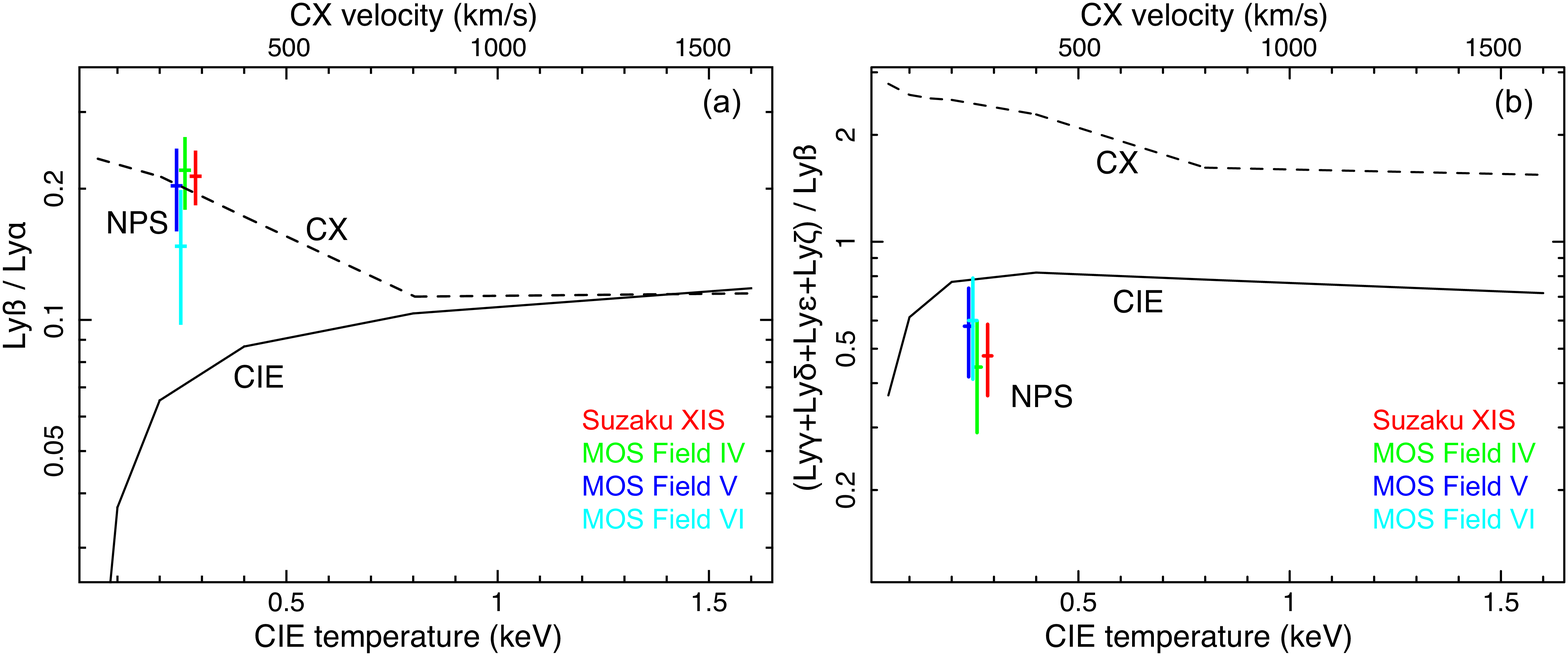}}
\caption{(a) The observed \oviii Ly$\beta$/Ly$\alpha$ ratios
  obtained with the XIS (red), the MOS field IV (green), field V (blue), and field VI (cyan) plotted against the best-fit temperatures. The solid line is
  the Ly$\beta$/Ly$\alpha$ versus temperature curve calculated by the {\tt CIE} model, and the dashed line shows the Ly$\beta$/Ly$\alpha$ versus collision velocity
  by the {\tt CX} model. (b) Same as panel a, but for the \oviii (Ly$\gamma$ + Ly$\delta$ + Ly$\epsilon$ + Ly$\zeta$)/Ly$\beta$ ratio. 
\vspace{0.5cm}
}
\label{lbla_fig}
\end{figure*}
%============================

In some objects, the enhanced forbidden-to-resonance ratios of triplet transitions might be explained by charge exchange recombination created by highly ionized
particles colliding with neutrals (e.g., Dennerl 2010). To examine such a possible component, we included a newly developed
{\tt CX} model (Gu et al. 2016) in the spectral fitting. This model calculates charge exchange emission by incorporating a set of velocity-dependent
reaction rates, followed by a radiative cascade calculation up to the atomic shell with principle quantum number $n=16$. In the fitting, the relative
velocity between hot and cold particles was left free to vary,
while the ionization temperature and metal abundances of the hot plasma were tied to those of the {\tt CIE} component. As plotted in Figure 2a,
the XIS fitting is slightly improved ($\Delta$C-statistics = 71.2 for a degree of freedom = 715), as the \ovii and \neix triplets are better reproduced by the recombining component. However, the
\oviii Ly$\beta$ deficiency cannot be solved with the {\tt CX} model. Setting the ionization temperature and metal abundances as free parameters
has a negligible effect on the fitting. For the MOS spectra, the {\tt CIE} + {\tt CX} model does not improve the fitting significantly over
the {\tt CIE} model alone.

Since the observed \oviii Lyman series cannot be described well under the current vision, we measure the \oviii line ratios and compare them directly
with the model. This is achieved by ignoring the \oviii ion during model calculation, and replacing it by putting six delta functions at the energies of
its Ly$\alpha$ (0.653 keV), Ly$\beta$ (0.774 keV), Ly$\gamma$ (0.817 keV), Ly$\delta$ (0.837 keV), Ly$\epsilon$ (0.847 keV), and Ly$\zeta$
(0.854 keV). We focus on two types of line ratios, the Ly$\beta$/Ly$\alpha$, and the (Ly$\gamma$ + Ly$\delta$ + Ly$\epsilon$ + Ly$\zeta$)/Ly$\beta$.
As shown in Figure 3a, the observed Ly$\beta$/Ly$\alpha$ ratios are higher, by a factor of $2-3$ from different instruments, than the value predicted by thin thermal
{\tt CIE} model with a balance temperature below 1.6 keV. The high Ly$\beta$/Ly$\alpha$ can be achieved by the {\tt CX} model, if the collision
velocity is lower than about 300 km s$^{-1}$. This result is consistent with those reported in Cumbee et al. (2016), which shows that the
high Ly$\beta$/Ly$\alpha$ ratio of \nex observed in M82 can be explained by their charge exchange code. On the contrary, Figure 3b shows that the observed (Ly$\gamma$ + Ly$\delta$ + Ly$\epsilon$ + Ly$\zeta$)/Ly$\beta$ ratios
agree better with the {\tt CIE} value; the {\tt CX} value is consistently higher than the observation by a factor of $3-5$ for the considered 
velocity range. This is probably due to the fact that the electrons captured in the O$^{8+}+$ H charge exchange would mainly fall onto
the high Rydberg states with $n = 4 - 6$, producing strong Ly$\gamma$ to Ly$\epsilon$ lines (e.g., Mullen et al. 2016). The $n = 3$ shell could
only be occupied by radiative cascade, and thus Ly$\beta$ line is weaker than the combined transitions from the shells with higher $n$. The line measurement
indicates that neither a pure {\tt CIE} or {\tt CX}, nor a combined model, can explain the observed NPS spectra.

%=========================%
\subsection{Additional ionized absorption}
%=========================%
%============================
%  FIG: ISM-CIE
%
\begin{figure*}[!]
\centering
\resizebox{0.8\hsize}{!}{\hspace{-1cm}\includegraphics[angle=0]{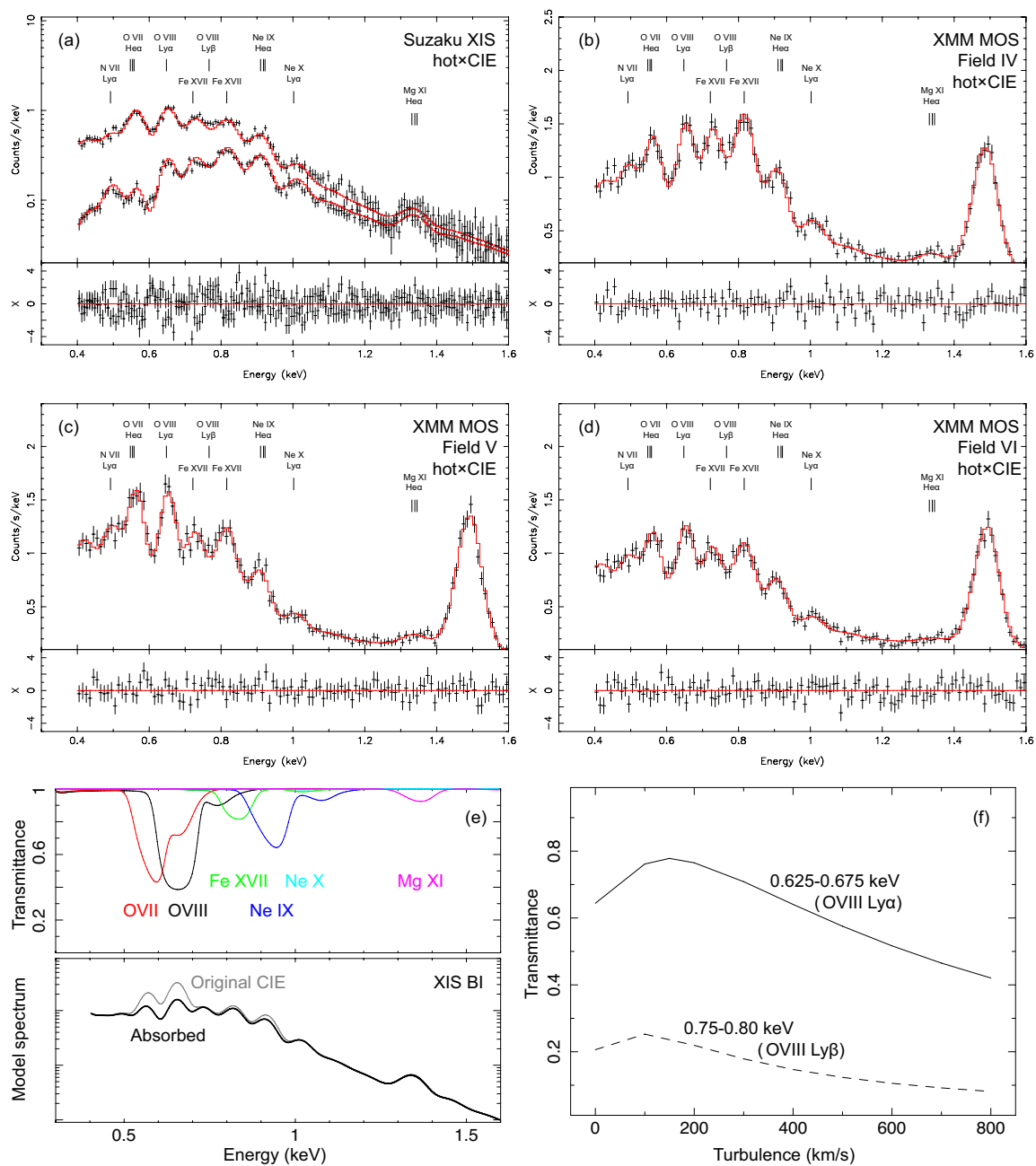}}
\caption{Same as Figure 1, but the best-fit models are one {\tt CIE} component subject to both neutral
  and ionized absorptions (a$-$d). Panel e shows the best-fit \ovii, \oviii, \fexvii, \neix, \nex, and \mgxi transmittances (upper),
  and the best-fit {\tt CIE} emission absorbed by the ionized particles (lower). The model is convolved with the {\it Suzaku} XIS1 response.
  Panel f plots the effective transmittances in the near-\oviii Ly$\alpha$ ($0.625-0.675$ keV; solid line) and near-\oviii Ly$\beta$ ($0.75-0.80$ keV;
  dashed line) bands as a function of turbulent velocity.
\vspace{0.5cm}
}
\label{CIE_fig}
\end{figure*}
%============================

The anomalous Ly$\beta$-to-Ly$\alpha$ ratio leads us to consider another scenario: the emission from the NPS might be obscured in the line of sight by an ionized absorber,
so that the transitions with large oscillator strengths, e.g., resonance lines of the triplets and Ly$\alpha$ lines, are partially absorbed. We 
utilized the {\tt hot} model in {\tt SPEX} to calculate the ionized absorption. For a given temperature, it calculates the ionization concentration for each 
ion, and then determines the cross section based on the oscillator strength, the thermal, and turbulent broadening. The plasma transmission is 
calculated by combining all the ionic transmissions. The ionized absorption was applied, in addition to the neutral absorption, to a {\tt CIE} component
describing the NPS emission, as well as to the CXB and GH components.  
In the fitting, the column density and temperature of the ionized absorber are set free to vary. The average systematic velocity and turbulent broadening 
of the absorber cannot be determined well by
the current data, and are always consistent with zero within error ranges. Thus we assume the absorber to be nearby and turbulence-free. For a similar reason, the metal
abundances of the absorber are fixed
to the solar value. As shown in Figures 4a$-$4d, the new model apparently better reproduces the \oviii Ly$\beta$ and \ovii/\neix triplets
for all XIS and MOS spectra. The best-fit C-statistics are 902.3 (degree of freedom = 714) for the XIS, and 130.5, 104.3, and 115.7 
(degrees of freedom $\approx 107$) for the three MOS data, significantly better than those obtained in \S3.1. The best-fit XIS model in Figure 4e
indicates that half of the emission in \oviii Ly$\alpha$ band is absorbed, while the Ly$\beta$ emission is absorbed by about 10\%.

As presented in Table 1, the ionized absorber has a best-fit temperature of $0.17-0.19$ keV, significantly lower than that of the NPS ($0.23-0.25$ keV).
The best-fit column densities by different instruments are in the range of $2.7-3.8 \times 10^{19}$ cm$^{-2}$. The resulting O abundance of the 
NPS becomes $0.6-0.7$ $Z_{\odot}$, more than twice as the value obtained in \S3.1. The other elements are less affected by the new model, probably because they have smaller
ionization concentrations than O at the temperature
of the absorber. The N/O abundance ratio becomes $0.68\pm0.22$, significantly lower than the one reported in M08 (N/O = 4).

We further investigate the effect of turbulence on the absorption feature in the \oviii bands. The opacity drops with the increasing turbulence, while the absorption
  line becomes broadened by turbulence so that it would affect also the adjacent continuum. As shown in Figure 4f, the effective transmittances in near-\oviii Ly$\alpha$ ($0.625-0.675$ keV)
  and near-\oviii Ly$\beta$ ($0.75-0.80$ keV) bands are plotted as a function of turbulent velocity. The calculation is based on the best-fit model for
  the XIS data, with only the \oviii lines and the continuum included. The obtained transmittances slightly differ from those shown in Figure 4e
  in which the model is folded with the XIS response. In both energy bands, the absorption feature reaches its maximum at a turbulence of $100-200$ km s$^{-1}$, and drops towards lower and
  higher velocities. As a result, for a turbulent absorber, the column density to fit the observed spectra would be lower by $\sim 10-20$\% with respect 
  to the static value at a velocity of 150 km s$^{-1}$,
  and becomes almost double at 1000 km s$^{-1}$.

Next we examine the XIS spectra for the possible charge exchange component. The emission model consists of a {\tt CIE} and a {\tt CX} components, both
are subject to the ionized and neutral absorption. The two emission components have the same temperature and metal abundances, while the collision velocity of
the {\tt CX} component is left free. As shown in Figure 2b and Table 1, the new model provides a minor improvement to the {\tt CIE}-alone fitting, with best-fit C-statistics of 877.2
for a degree of freedom of 712. Based on the best-fit $\chi^{2}$ values, an $F$-test shows that the {\tt CX} component is significant on the $>3\sigma$
confidence level. The charge exchange emission contributes $\approx 8$\% of the entire NPS emission in $0.3-1.6$ keV. The best-fit collision velocity is
$150^{+150}_{-100}$ km s$^{-1}$. As seen in Table 1, the {\tt CIE} temperature remains intact, while the abundances are slightly affected by including the {\tt CX}.
The best-fit N/O ratio becomes $0.75\pm0.29$. The new fitting further prefers a higher column density of the ionized absorption, $5.7\pm 1.3 \times 10^{19}$ cm$^{-2}$, to balance out the additional line emission introduced by {\tt CX} in the spectra.

\section{Discussion}
%============================
%  FIG: ISM-CIE
%
\begin{figure*}[!]
\centering
\resizebox{0.8\hsize}{!}{\hspace{-1cm}\includegraphics[angle=0]{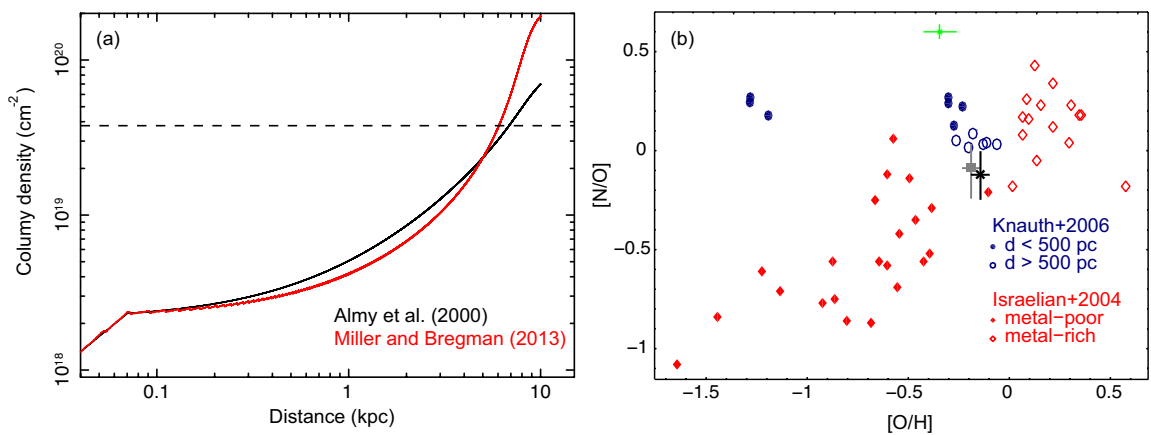}}
\caption{(a) Integrated Earth-centered ISM column density profiles based on the models reported in Almy et al. (2000, black) and Miller \& Bregman (2013, red). Extra
  column density from the local hot bubble is included in the central 70 pc.
  The dashed line shows the observed value based on {\tt CIE} model with ionized absorption. (b) The [N/O] verseus [O/H] diagram. The best-fit NPS results obtained
  with {\tt CIE} and {\tt CIE} plus {\tt CX} models with the new absorptions are shown in black and grey error bars, respectively. The M08 result is plotted in green.
  The abundance patterns of the cold ISM based on local star and distant star absorption measurements by Knauth et al. (2006) are plotted in blue filled circles and blue
  empty circles, respectively. The metal-poor and metal-rich Galactic halo star data from Israelian et al. (2004) are marked by red filled diamonds and red empty diamonds, respectively. 
\vspace{0.5cm}
}
\label{CIE_fig}
\end{figure*}
%============================

By revisiting the {\it Suzaku} and {\it XMM-Newton} data of the NPS region, we discovered that the soft X-ray spectra can be well described
by a single-phase thermal component, with a temperature of $0.23-0.25$ keV, absorbed by at least two species of foreground materials, in both neutral and ionized states.
The key evidence for the ionized absorber is the unusually high \oviii Ly$\beta$ line relative to other Lyman series. Assuming a nearby turbulent-free plasma, the
hot absorber exhibits a balance temperature of $0.17-0.20$ keV and column density of $\sim 3-5 \times 10^{19}$ cm$^{-2}$. 
A charge exchange component is
marginally detected only with the {\it Suzaku} XIS data. The oxygen abundance of the NPS is then obtained to be $0.6-0.7$ $Z_{\odot}$, apparently
higher than those reported in W03 and M08. The Fe/O ratio is consistent with the solar values within measurement uncertainties, while the 
N/O becomes slightly sub-solar.

Next we shed light on the origin of the ionized absorber based on the derived properties. As shown in \S 3.2, the balance temperature of the
absorber is $0.17-0.20$ keV, lower than the NPS plasma temperature $\approx 0.23-0.25$ keV on $> 90$\% confidence level. This means that it
cannot be fully ascribed to the self-absorption of the NPS plasma. On the other hand, the local hot bubble alone cannot be the absorber either.
The temperature of the local hot bubble ($\approx 0.1$ keV) is lower than the observed value, and assuming a line-of-sight scale
of $40-90$ pc and density of $0.01$ cm$^{-3}$ (e.g., W03), the column density is estimated to be $1.2-2.7 \times 10^{18}$ cm$^{-2}$, accounting
$<10$\% of the obtained value (\S3.2).

The obtained properties of the ionized absorber are consistent with those reported in the absorption studies on Galactic compact object 4U 1820$-$303
(Hagihara et al. 2011), and extragalactic objects PKS 2155$-$304 (Hagihara et al. 2010), LMC X$-$3 (Yao et al. 2009), and Mrk 509 (Pinto et al. 2012). This leads us to the scenario
that the Galactic ISM contributes significantly to the observed ionized absorption. The temperature $\approx 0.2$ keV appears to be self-consistent with 
that of the Galactic ISM included as a background component in \S2.2. It also agrees well with previous measurements of the ISM temperature in the
Galactic halo (e.g, Smith et al. 2007) and Galactic bulge (e.g., Almy et al. 2000). To calculate the ISM absorption column density, we employ the 3-D ISM density models
from Almy et al. (2000, in their Fig. 5) and Miller \& Bregman (2013, ``spherical-saturated'' model in their Table 2). The former is based on {\it ROSAT}
3/4 keV observation of the emission in the Galactic bulge region, while the latter is focused on the Galactic halo, and utilized line absorption measurement
on background objects. The two models provide galactocentric ISM density profiles, which are then transformed into Earth-centered line-of-sight distance profiles
by using Eqs.~1$-$3 of Miller \& Bregman (2013). The sky coordinate of the {\it Suzaku} pointing is used in the center transformation. By integrating the density
over distance, we calculate the column density distance profiles and present them in Figure 5a. It shows that the two ISM models are roughly consistent with each
other. Let us consider an extreme case, in which the ionized absorption is fully due to the Galactic ISM, the models predict that the part of NPS covered by the XIS would be at a 
distance of $\sim 6-7$ kpc. This is still well in line with the recent measurements of the NPS distance with radio data by Sun et al. (2014) and Sofue (2015). 

%The intervening ISM would not only absorb portion of the NPS emission, but also emit itself thermal radiation at the same temperature $\approx 0.2$ keV.
%To estimate the ISM flux, we employ again the density model of Almy et al. (2000). By integrating the Earth-centered density profile (Fig. 5a)
%to a halo scale of 20 kpc (e.g., Miller \& Bregman 2013), the ISM emission measure is determined, and is thus converted to a total {\it Suzaku} XIS rate
%of 0.18 count s$^{-1}$ in $0.3-1.6$ keV. This contributes to a small fraction $\approx 12$\% of the NPS emission, part of which has already been accounted by the
%GH component described in \S2.2. The extra ISM component might pose an additional uncertainty of $10-20$\% in the \ovii line measurement, but it cannot affect
%our main conclusion, which is based on mostly the \oviii Lyman series.

As described in \S3.2, the over-solar N/O abundance ratio reported in M08 can be migrated to the large opacity of \oviii Ly$\alpha$ line.
In Figure 5b, the new results are plotted in a [N/O] verseus [O/H] diagram, and
are compared with the abundances of Galactic halo stars measured in Israelian et al. (2004). The NPS values are consistent with the implied Galactic stellar evolution by 
lying in the gap between the ``metal-poor'' and ``metal-rich'' subsamples of stars. At the same time, the abundance patterns of Galactic cold ISM, based on {\it HST} and {\it FUSE}
observations of \oi and \nni absorptions against stars (Knauth et al. 2006), are also plotted in the same diagram of Figure 5b. Despite of the large uncertainties, the NPS results appear to agree
better with the abundance patterns of the distant ISM
($d > 500$ pc), than with those of the local ISM ($d < 500$ pc). This also supports the scenario that the NPS is a structure in the Galactic halo rather than in the solar
neighborhood.

\section{Summary}

By re-analyzing the {\it Suzaku} and {\it XMM-Newton} data of the North Polar Spur, we detected an anomalously high \oviii Ly$\beta$ line relative to other Lyman series
in four different fields. It prefers an ionized absorption model over a charge exchange component, which suggests that the NPS is partly obscured by foreground
plasma, presumably Galactic hot ISM, with a temperature of $0.17-0.20$ keV and column density of $3-5 \times 10^{19}$ cm$^{-2}$. After correcting the absorption, the
oxygen abundance of the NPS changes from $\sim 0.2$ $Z_{\odot}$ to $\sim 0.7$ $Z_{\odot}$, and the abundance ratio between nitrogen and oxygen becomes even a bit lower than the solar value. Combining the absorption and abundance measurements, it is suggested that the North Polar Spur is likely an object in the Galactic halo, supporting the ``Galactic center origin'' scenario.
This exercise provides a good example to show that an accurate spectral model is crucial to ensure reliable scientific output.

\section*{Acknowledgments}
We thank the referee, Randall Smith, for making valuable comments on the manuscript. SRON is supported financially by NWO, the Netherlands Organization for Scientific Research.

\appendix

%====================%
\section{Possible issues with the {\it Suzaku} data calibration}
%====================%

%============================
%  FIG: BD+30
%
\begin{figure*}[!]
\centering
\resizebox{0.8\hsize}{!}{\hspace{-1cm}\includegraphics[angle=0]{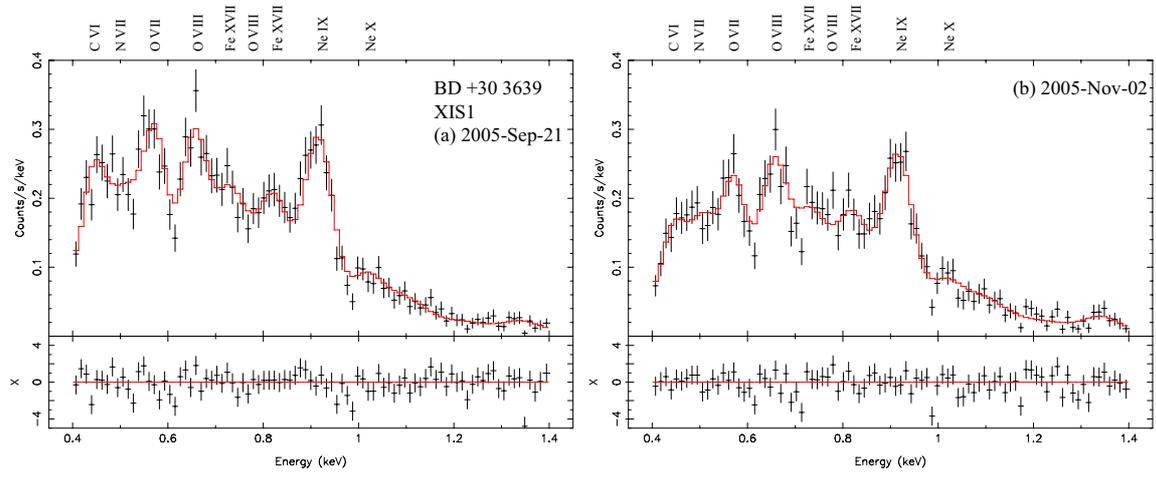}}
\caption{The XIS1 spectra of BD +30 3639 taken before the NPS (a) and after the NPS (b), modeled with a single-phase {\tt CIE} component
  absorbed by Galactic neutral material. 
\vspace{0.5cm}
}
\label{BD+30_fig}
\end{figure*}
%============================

As reported in M08, the {\it Suzaku} NPS data might be affected by a build-up of contamination on the optical blocking filter in the
early phase of the mission. The effective area of the XIS around 0.5 keV is therefore dependent on time and chip location. To examine
the possible calibration uncertainty, we analyzed two {\it Suzaku} datasets observed around the same time as the NPS data, and compared
their spectra in the $0.4-1.4$ keV band. The two observations were made in 2005 September 21 and 2005 November 2, which pointed to a common
target, planetary nebula BD +30 3639. The same data were reported in Murashima et al. (2006). 

The XIS data of BD +30 3639 were screened in the same way as described in \S 2.1. The source spectra were taken from the central 3$^{\prime}$, and
the background region was defined as a surrounding annulus with outer radius of 5$^{\prime}$. Following \S 2.2, we modeled the background spectrum
and corrected it in the source spectra. The two data were fit simultaneously. As shown in Figure A.1, the two source spectra are nicely fitted by
a {\tt CIE} model absorbed by Galactic neutral material. The best-fit plasma temperature (0.18 keV) and abundances agree well with those reported
in Murashima et al. (2006). The spectrum below 1 keV appears to be more absorbed in the second observation,
probably due to the contamination on the optical blocking filter. This effect has been fully corrected by the response files. This exercise proves
that the spectral features on \ovii, \oviii, and \neix discovered in the NPS data (\S3.1) are unlikely due to calibration issues.

Similar to the NPS, the BD +30 3639 spectra also show two prominent \fexvii lines around, or partially blended with the \oviii Ly$\beta$ line. While
the NPS spectra exhibit a strong excess in the \oviii Ly$\beta$ line (\S3.1), the same energy band in the BD +30 3639 spectra is well-fitted 
by the thermal model. This further indicates that the \oviii Ly$\beta$ excess in the NPS is unlikely to be an artifact by either \fexvii blending, or poor atomic model calculation of the \fexvii + \oviii complex.

\end{document}